\documentclass[acmsmall,authorversion,nonacm]{acmart}
\AtBeginDocument{%
  }

\setcopyright{acmlicensed}
\copyrightyear{2018}
\acmYear{2018}
\acmDOI{XXXXXXX.XXXXXXX}
\acmConference[Conference acronym 'XX]{Make sure to enter the correct
  conference title from your rights confirmation email}{June 03--05,
  2018}{Woodstock, NY}
\acmISBN{978-1-4503-XXXX-X/2018/06}

\usepackage{multirow}
\usepackage{tablefootnote}

\begin{document}

\title{Understanding and Supporting Co-viewing Comedy in Virtual Reality with Embodied Expressive Avatars}

\author{Ryo Ohara}
\affiliation{%
  \institution{The University of Tokyo}
    \city{Tokyo}
  \country{Japan}}
\email{rohara@cyber.t.u-tokyo.ac.jp}

\author{Chi-Lan Yang}
\affiliation{%
  \institution{The University of Tokyo}
    \city{Tokyo}
  \country{Japan}}
\email{chilan.yang@cyber.t.u-tokyo.ac.jp}

\author{Takuji Narumi}
\affiliation{%
  \institution{The University of Tokyo}
    \city{Tokyo}
  \country{Japan}}
\email{narumi@cyber.t.u-tokyo.ac.jp}

\author{Hideaki Kuzuoka}
\affiliation{%
  \institution{The University of Tokyo}
    \city{Tokyo}
  \country{Japan}}
\email{kuzuoka@cyber.t.u-tokyo.ac.jp}

\renewcommand{\shortauthors}{Ohara et al.}

\begin{abstract}
Co-viewing videos with family and friends remotely has become prevalent with the support of communication channels such as text messaging or real-time voice chat.
However, current co-viewing platforms often lack visible embodied cues, such as body movements and facial expressions. This absence can reduce emotional engagement and the sense of co-presence when people are watching together remotely. Although virtual reality (VR) is an emerging technology that allows individuals to participate in various social activities while embodied as avatars, we still do not fully understand how this embodiment in VR affects co-viewing experiences, particularly in terms of engagement, emotional contagion, and expressive norms.
In a controlled experiment involving eight triads of three participants each (N=24), we compared the participants' perceptions and reactions while watching comedy in VR using embodied expressive avatars that displayed visible laughter cues. This was contrasted with a control condition where no such embodied expressions were presented.
With a mixed-method analysis, we found that embodied laughter cues shifted participants’ engagement from individual immersion to socially coordinated participation.
Participants reported heightened self-awareness of emotional expression, greater emotional contagion, and the development of expressive norms surrounding co-viewers' laughter.
The result highlighted the tension between individual engagement and interpersonal emotional accommodation when co-viewing with embodied expressive avatars.
We discuss design implications for immersive media systems that seek to balance personal engagement with socially shared emotional experience.
\end{abstract}

\ccsdesc[500]{Human-centered computing~Empirical studies in collaborative and social computing}


\keywords{Co-viewing Experience, Virtual Reality, Avatar-mediated Communication, Emotional Contagion}

\received{20 February 2007}
\received[revised]{12 March 2009}
\received[accepted]{5 June 2009}

\begin{teaserfigure}
    \centering
    \includegraphics[width=\textwidth]{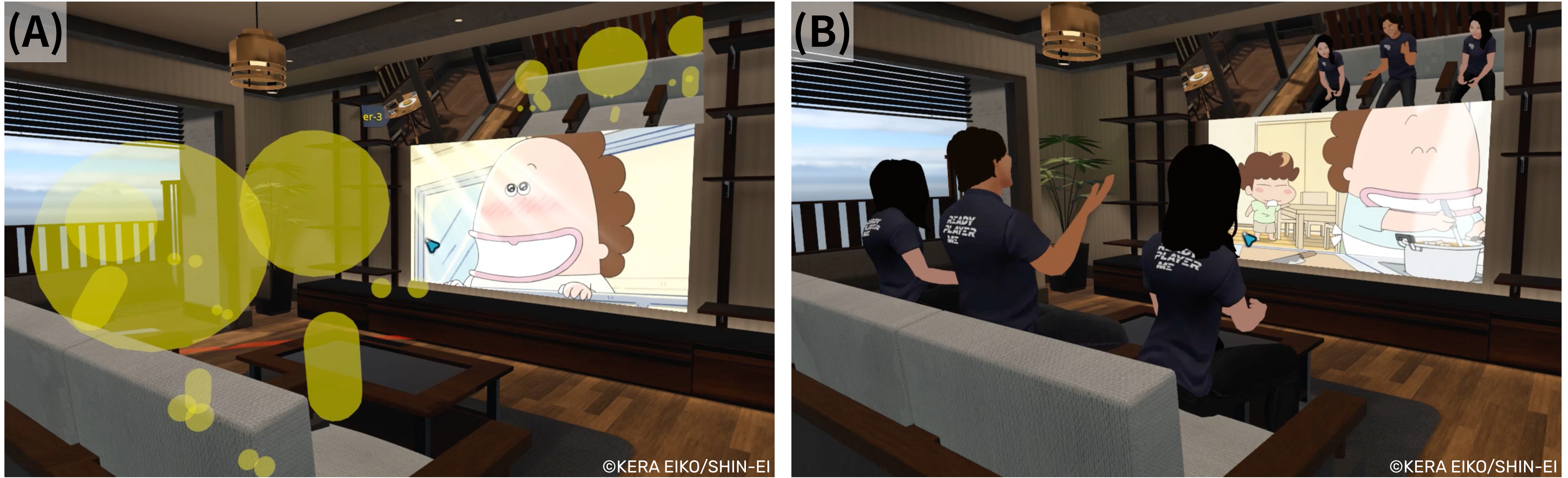}
    \caption{Co-viewing comedy in VR under two conditions. (A) \textit{Voice Only (VO)}: participants communicated solely via voice chat and were not represented by avatars. Yellow translucent figures are added for illustrative purposes and were not visible during the actual study. (B) \textit{Voice + Laughter Cues (VLC)}: participants were embodied in full-body avatars capable of displaying facial expressions and whole-body laughter animations.}
\label{fig:teaser}
\end{teaserfigure}

\maketitle

\section{INTRODUCTION}
Watching online content has transformed from an individual activity into a socially embedded experience.
Contemporary video platforms increasingly support co-viewing through interactive features such as text-based chat and real-time reactions \cite{VideoSharingSurvey}.
One widely studied example is Danmaku, a feature that overlays time-synced viewer comments directly onto video content. Prior work has shown that such features could foster engagement~\cite{EngagementOnLiveStreaming}, emotional contagion~\cite{EmotionalContagionInLiveStreaming}, and a sense of social co-presence~\cite{DanmakuIntJHumComputInteract, SocialPresenceDanmaku, DanmakuCSCW} among distributed co-viewers.
Meanwhile, voice-based co-viewing platforms such as Discord Stage Channels\footnote{Discord Stage Channels: \url{https://discord.com/stages}} and Apple SharePlay\footnote{Share Play: \url{https://www.apple.com/newsroom/2021/11/shareplay-powers-new-ways-to-stay-connected-and-share-experiences-in-facetime/}} have emerged, enabling close social ties to converse during media consumption~\cite{SharedVideoExperiences,SmartTVInterface}.
As a result, co-viewing now spans multiple modalities, including textual, auditory, and visual.

To deepen this experience, virtual reality (VR) has become an emerging platform for immersive co-viewing.
Recent systems have introduced autonomous avatars to enhance the sense of social presence for co-viewing in VR~\cite{VRCoviewing4}, detect humorous moments, and generate laughter from surrounding avatars~\cite{EmbodiedLaughTrack}, or animate user-generated Danmaku into nonverbal avatar gestures~\cite{DanmakuAvatar}.
While these systems enhance individual engagement through social signals, they largely focus on the effect of simulated or pre-scripted behaviors on individual experience, rather than interpersonal interactions in co-viewing.
Importantly, we lack understanding about how users exchange and regulate emotions through embodied, mutual expressions during co-viewing.

This presents a significant gap: while prior work has examined co-viewing as a mediated social activity, little is known about how embodied expressive cues, such as avatar gesture and facial expression, shape shared media experiences in immersive environments. As social VR platforms like VRChat\footnote{VRChat: \url{https://hello.vrchat.com}} become increasingly popular for engaging in various social events~\cite{VRGoesSocial}, understanding how these embodied interactions influence engagement, emotional contagion, and collective experience is important.
In particular, there is a need to investigate the tensions between focused content engagement and emotional expression from co-viewers, and how users actively negotiate these through co-viewing practices.
Our study addresses this gap by examining how co-viewers use embodied avatars to express affect and regulate emotional expression in shared VR settings.

We asked three research questions to address these issues:
\begin{itemize} 
    \item[\textbf{RQ1:}] \textit{How do embodied laughter cues influence individual \textbf{engagement} when co-viewing comedy in VR?}
    \item[\textbf{RQ2:}] \textit{How do embodied laughter cues shape \textbf{emotional contagion} between co-viewers when co-viewing comedy in VR?}
    \item[\textbf{RQ3:}] \textit{How do embodied laughter cues contribute to the \textbf{norm} of emotional expression within groups when co-viewing comedy in VR?}
\end{itemize}

Different from prior work that relies on system-controlled avatars or two-dimensional interfaces, this study examines how real-time, emotionally expressive interaction unfolds between users who co-view content through self-animated avatars in immersive three-dimensional environments.
Rather than simulating affect through automated behaviors, our approach focuses on user-driven expression, allowing participants to shape their co-viewing experience through embodied gestures and avatar-based cues.
We focus on comedy as the target genre for three key reasons.
First, entertainment, especially comedic content, remains one of the most prevalent contexts for co-viewing \cite{YouTubeAnalysis,ChallengesOnCowatchingYouTube}, as features like Danmaku are most frequently used at humorous moments \cite{DanmakuCSCW}.
Second, laughter and related affective expressions are inherently social and contagious \cite{ContagiousLaughter,AntiphonalLaughter, NotAllLaughsAreAlike, ReconsideringLaughter}, making them particularly well-suited for studying emotional exchange.
Third, comedy has been widely adopted in prior co-viewing research \cite{EmbodiedLaughTrack,AnalysisofCoviewingwithVA,HumorwithOutgroups}, providing a familiar and comparative baseline for this study. 

To examine how embodied cues influence emotional dynamics in co-viewing, we conducted a controlled mixed-methods study with eight triads (N=24), comparing voice-only communication with avatar-mediated co-viewing that visualized embodied body movement and laughter.
Our findings show that expressive avatars amplified emotional contagion, shifted attention toward social interaction, and fostered shared norms around expressive emotional cues.
While many participants reported heightened engagement and emotional alignment, others expressed a desire for greater ambiguity and interpersonal distance, depending on content and group composition.
These findings highlight the need for co-viewing systems to support flexible emotional expression and reception, designed to accommodate different social configurations and types of content.
This study advances our understanding of how embodied affective cues support emotional coordination and regulation in co-viewing in VR.
We discuss design implications for avatar-mediated interaction in social VR platforms that aim to balance individual engagement with interpersonal co-viewing experiences.

\section{RELATED WORK}
\subsection{Individual and Social Aspects of Video Consumption}
Enriching individual and social video consumption using interactive technology has been extensively explored in the HCI field.
On the individual aspect, studies have investigated how interactive features, such as enabling users to share emotional responses~\cite{SocialTV, InterpersonalBiasInTV, wang2012TVwatching, tu2016coviewingRoom}, actively interact with video content~\cite{MultimodalInteractionInLiveStreamingForLanguageLearning,ChatHasNoChill} or adding danmaku~\cite{DanmakuCSCW,DanmakuIntJHumComputInteract,SocialPresenceDanmaku}, can enhance viewers' engagement, comprehension, and emotional experience during media viewing.
On the social aspect, co-viewing platforms such as Twitch and Bilibili have been studied for how they mediate emotional sharing and social presence through textual, audio, and visual channels~\cite{joden2022building, DanmakuTransSoc}.
Related works showed that introducing virtual characters during sports viewing activities~\cite{CowatchingWithAgents} or projecting audience silhouettes~\cite{AudienceSilhouettes} could enhance the social presence of distributed co-viewers.
While these systems foster social interaction, they typically rely on disembodied forms of communication. Virtual reality (VR) environments provide an effective platform for embodied interaction, enabling users to convey emotions through gestures and enhanced facial expressions~\cite{Embodiement2}, while also potentially experiencing a heightened sense of social presence with co-viewers.

The Proteus effect~\cite{ProteusEffect} describes the phenomenon in which individuals' behaviors and self-perceptions conform to the characteristics of their digital avatars. In VR, this effect has been shown to influence users’ perception and behaviors, including confidence and conformity~\cite{yee2009proteus}. However, we still lack a clear understanding of how embodied expressive avatars affect people's emotional expression and perception during co-viewing in VR. On one hand, it is possible that embodied cues enhance social presence, thus leading to an engaged co-viewing experience. 
Based on the Proteus effect~\cite{ProteusEffect, yee2009proteus}, when co-viewing comedic content using emotionally expressive avatars, being embodied in expressive avatars may shape users’ perceived expressiveness and willingness to engage in visible affective reactions, such as laughter or gestural reactions. Embodying an avatar designed to convey emotional cues (e.g., smiling, laughing, or exaggerated movements) can lead users to internalize these expressive traits, thereby amplifying their emotional responses.
On the other hand, it is also possible that viewing co-viewers' embodied cues may shift attention away from the content and toward performative social dynamics, thus reducing the engagement during co-viewing. Understanding the relationship between individual engagement and interpersonal emotional coordination in co-viewing in VR is crucial for designing embodied avatars that support authentic and comfortable emotional expression in shared virtual experiences.

\subsection{Emotional Contagion in Computer-Mediated Interaction}

Emotional contagion is the phenomenon where individuals adopt others’ emotions through implicit or explicit cues, often without conscious awareness~\cite{EmotionContagion,EmotionContagion2}. In face-to-face interactions, emotional contagion occurs through vocal tone, facial expressions, body language, and other nonverbal behaviors~\cite{hatfield2014new}. In computer-mediated communication (CMC), while the nonverbal cues are more limited, studies have shown that emotional contagion still emerges through mechanisms such as textual emotive expressions, emojis, linguistic mimicry, and synchronous behavior~\cite{EmotionCognitionSNS1,EmotionCognitionSNS2,EmotionCognitionSNS3,EmotionContagionSNS4}. Emotional contagion can influence not only individuals' emotional states but also group dynamics, contributing to emotional alignment in collaborative and social contexts~\cite{EmotionalContagionInCMC,EmotionalContagionInLiveStreaming}.

In group interactions online, emotional contagion can influence collective mood, influencing engagement levels, trust formation, and group cohesion. For example, \citeauthor{barsade2002ripple} demonstrated that positive emotional contagion in workgroups leads to greater cooperation, task coordination, and overall team performance~\cite{barsade2002ripple}. Similarly, \citeauthor{EmotionalContagionInCMC} found that in text-based collaborative environments, the negative emotional tone of one participant could be sensed by their remote counterpart and significantly reduced the positive affect of others~\cite{EmotionalContagionInCMC}.

Despite the rich understanding of emotional contagion in text-based interactions, we know little about whether and how it occurs in interactions mediated by avatars. Avatars can display various nonverbal cues, such as gaze, gestures, and facial expressions, which could facilitate effective emotional contagion. However, evidence suggests that emotional contagion does not consistently happen in avatar-mediated interactions~\cite{van2024emotion}. As VR becomes a platform for co-viewing, it is essential to clarify whether and how being represented by avatars can lead to emotional contagion and how this influences the relationships among co-viewers.



\subsection{Shared Social Experience using Embodied Avatars in Virtual Reality}

Social VR platforms such as VRChat allow users in different locations to participate in a variety of shared activities~\cite{VRGoesSocial, VRCollabo1}. Users can play games, have meetings with colleagues, study with friends, interact with strangers online~\cite{SocialVREmbodiement}, and drink with other users~\cite{chen2024d}. The embodied presence of users enhances their self-awareness and their awareness of others within the social VR environment
~\cite{SocialVREmbodiement}.

Across these diverse activities, users often create shared expectations or unwritten community standards that influence their behaviors, self-presentation, and interactions on the platform~\cite{hide, SocialVRAvatar}. The norms that emerge can be shaped by the affordances of the platform. For instance, in VRChat, users tend to engage in more creative and dramatic self-presentation due to the platform's emphasis on avatar customization and distinctive virtual environments~\cite{VRChatGenderExpression, SocialVRAvatar}. In contrast, AltspaceVR promotes social environments that resemble real-life interactions, encouraging users to present themselves in ways that align with their physical identities~\cite{AltspaceVRContact, SocialVRAvatar}.

Although previous research has provided insights into how users establish norms in social VR environments, particularly concerning self-presentation~\cite{SocialVRAvatar} and the safeguarding of personal spaces~\cite{PersonalSpaceSocialVR}, there is limited understanding of how these norms change during emotionally intense shared experiences, like watching comedy together or attending large sports events. In particular, the role of expressive embodied avatars in shaping emotional expression and interpersonal coordination during video co-viewing remains underexplored. As social VR becomes an emerging channel for watching videos together~\cite{VRScreenViewing, VRCoviewing4}, it is critical to understand how avatar design influences not only individual affective experiences but also the social norms around emotional expression, reception, and mutual attention. 
Earlier studies have shown that co-viewing with avatars controlled by computers can enhance individuals' sense of co-presence and engagement~\cite{VRCoviewing4}. However, it is still unclear how co-viewing with embodied avatars, representing the co-viewers themselves, affects the mutual expression of emotions and emotional contagion between them.
Thus, this work addresses this gap by investigating how being embodied in expressive avatars in social VR influences the development of co-viewing norms and emotional coordination among co-viewers.

\section{METHOD}
We conducted a controlled experiment in an immersive VR environment to investigate how embodied laughter cues shape remote co-viewing experiences.
Specifically, we examined how these cues influence individual engagement during co-viewing sessions (RQ1), facilitate emotional contagion between co-viewers (RQ2), and contribute to the formation of shared norms for emotional expression within co-viewing groups (RQ3).

\subsection{Experiment Design}
To examine the impact of visible laughter cues in VR-based remote co-viewing, we conducted a within-subjects experiment by comparing two conditions, \textbf{Voice Only (VO)} and \textbf{Voice + Laughter Cues (VLC)}.
In \textbf{VO} condition, participants viewed comedy videos together using real-time voice chat only, with no visible laughter expressions (\autoref{fig:teaser} (A)).
In \textbf{VLC} condition, in addition to voice chat, participants were able to observe embodied laughter cues shared among group members (\autoref{fig:teaser} (B)).
These cues were controlled by the participants, who could trigger avatar animations representing laughter at moments of their choosing.
We did not include a condition with idling avatars, as prior studies have shown that mismatches between auditory and visual cues can undermine social presence~\cite{AppearanceAndBehavior}.
This effect is particularly pronounced for laughter, which relies on temporal synchrony and mutual responsiveness for emotional alignment~\cite{EmbodiedLaughTrack}.
In this context, \textbf{VO} condition serves as a baseline, aligning with popular video live-streaming platforms such as Discord Stage, which are frequently utilized for co-viewing.

Participants joined the study in triads and viewed four 7-minute episodes of Atashin'chi\footnote{Atashin'chi Official Channel \url{https://www.youtube.com/@Atashinchi}}, a Japanese animated series known for its accessibility across age groups and minimal cultural or linguistic barriers.
We set the group size of three based on prior work, which suggested that while four is the minimum size for optimal conversation~\cite{ConversationGroupSize}, smaller groups can share and experience laughter more effectively in such a small setting~\cite{LaughterGroupSize}.
Participants were seated in a laboratory setting and separated to prevent any visual or auditory contact during the sessions.
To prevent demand characteristics and elicit spontaneous behavioral responses, participants were informed that the study aimed to identify humorous segments in video content, rather than revealing its actual focus on co-viewing dynamics.
Participants watched two different clips in each condition. 
The order of conditions was counterbalanced across groups to mitigate order effects.
In both conditions, participants could manually trigger laughter expressions of their avatar using the same interface.

\subsection{Overview of the VR Co-viewing System}
\subsubsection{Embodied Laughter Cues}
\begin{figure}
    \centering
    \includegraphics[width=\textwidth]{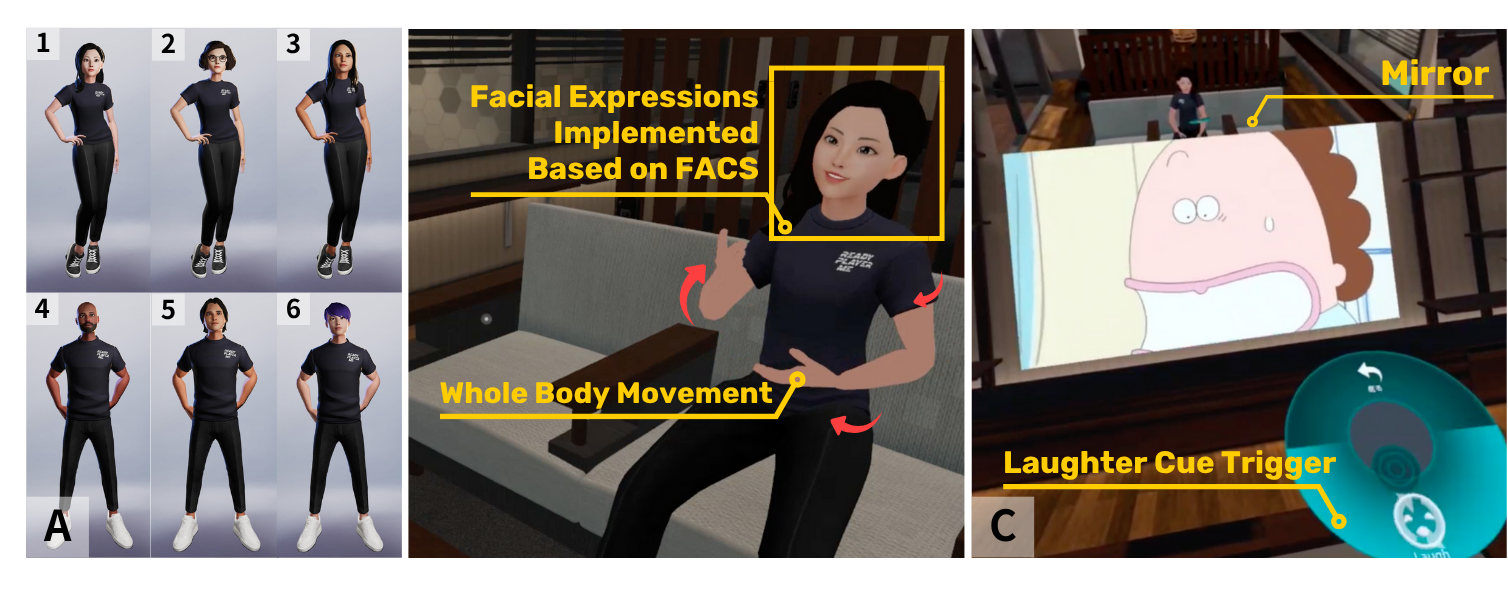}
    \caption{Implementation overview. (A) We provided six default avatars from ReadyPlayerMe for participants to choose from. (B) Expressive cues of the avatars include FACS-based facial expressions (e.g., cheek raising, mouth opening) and whole-body movements for seated laughter animation. (C) Users' first-person view in VR. The virtual co-viewing environment consists of a TV screen, a three-person sofa, and a mirror above the TV, allowing users to observe their own and co-viewers' expressions. Users can trigger the laughter expressions of their embodied avatars with the laughter cue trigger on the panel in front of them}
    \label{implementation}
\end{figure}
We implemented avatar-based laughter expressions as visibly expressed bodily behaviors rather than symbolic representations such as emojis.
For example, avatars displayed rhythmic upper-body movement and smiling facial expressions.
These visible embodied cues were modeled on natural human laughter and based on established designs from prior work~\cite{EmbodiedLaughTrack}.
Emotional contagion can be facilitated by visible motor mimicry~\cite{EmotionContagion,EmotionContagion2}, and immersive environments such as virtual reality may enhance this effect by promoting a sense of embodiment.

Prior research has described laughter as comprising facial expressions, body movements, and vocal sounds~\cite{SurveyOnLaughterDetectionMeasurementClassfication}.
In this study, we adopted only the visual components of facial and bodily expressions without making any changes to their voice because vocal laughter was transmitted naturally through voice chat.
The effectiveness of avatar-based visual laughter cues has been demonstrated in previous work, and our study extends this approach to a co-viewing context to examine their impact on emotional contagion~\cite{EmbodiedLaughTrack}.

\paragraph{Facial expressions of the avatar:}
Facial expressions were implemented using the Facial Action Coding System (FACS)~\cite{FACS}, a standardized framework for representing facial movements through Action Units (AUs).
This framework was selected due to its widespread adoption in avatar facial animation research\cite{FACSAvatar1, FACSAvatar2}.
Based on prior work on laughter expression~\cite{TheExpressivePatternOfLaughter,TowardsMultimodalExpressionOfLaughter,FACS25,FACS25_2}, we included AUs commonly associated with spontaneous laughter, such as AU6 (cheek raiser), AU12 (lip corner puller), and AU25 (lips part). 
Expressions were implemented using Character Creator 4\footnote{\url{https://www.reallusion.com/jp/character-creator/}}, which provides predefined mappings between AUs and avatar blend shapes.

\paragraph{Body movement of the avatar:}
For body movement, we reused the ``Sitting Clapping'' animation from Mixamo\footnote{Mixamo \url{https://www.mixamo.com/}}, which has been employed in previous work to represent seated human laughter.
The motion includes features such as torso sway and hand movement, and its reuse ensured compatibility with prior implementations and ecological validity~\cite{SurveyOnLaughterDetectionMeasurementClassfication,BodyMotion1,BodyMotion2}.
Avatar models were selected from the default ReadyPlayerMe set to reflect demographic diversity and were standardized to support identical expression capabilities (\autoref{implementation} (A)).

\subsubsection{Interaction Method and Co-Viewing Setup}
We built the virtual co-viewing environment using Unity and the VRChat SDK, and deployed it as a custom world on the social VR platform, VRChat.
This setup enabled multiple remote users to watch video content side-by-side, simulating the experience of co-presence in a shared physical space.
The environment was designed to resemble a typical living room, featuring a three-person sofa placed in front of a large television screen.
The screen was configured to play YouTube videos in sync across all participants.
To support mutual visibility of avatar reactions, a mirror was installed above the screen, allowing participants to see both their own and others’ facial expressions and body movements of their avatars.
This mirror-based design follows standard practices in virtual environments that utilize mirrors to enhance avatar embodiment~\cite{VRMirror1, VRMirror2}.
The implemented VRChat world is accessible via a dedicated link\footnote{URL omitted for anonymous review. The link to the implemented VRChat world will be provided upon publication.} to facilitate replication and further investigation.

The embodied laughter cues were manually triggered by participants using the default Action Menu in VRChat (\autoref{implementation} (C)).
This interface is commonly used for emotional expressions like Emotes in VRChat and can be easily operated by slightly tilting the controller stick.
We opted for manual input over automatic detection, as current laughter recognition technologies remain limited in accuracy, especially in multi-speaker and online voice-chat environments, and are often sensitive to individual and cultural differences~\cite{LaughterAutoDetect}.
Manual input provides participants with explicit control over when and how to express laughter.
It allows them to externalize their internal reactions, including subtle amusement that may not be vocalized, without relying on potentially unreliable detection.
This approach mirrors the way users selectively express emotion using emojis or reactions in video streaming and social platforms, and aligns with our research goals of studying intentional emotional sharing.
Additionally, prior work has shown a strong consistency between individuals’ self-reports of humor and their psychological sensitivity to humorous stimuli~\cite{Nomura}, supporting the appropriateness of self-initiated laughter cues in capturing participants’ subjective experiences.

\subsection{Participants}
As shown in Table \ref{tab:participant_info}, we recruited 24 participants, consisting of 17 males and 7 females, with an average age of 33.1 years ($SD$ = 17.3 years).
The sample size was determined by referencing related works that also focused on the co-viewing experience \cite{AmIInTheTheater,AgainTogether}.
In our qualitative analysis, we confirmed that no new codes or themes emerged in the later stages of coding, suggesting that theoretical saturation was reached.
An a priori power analysis with G*Power 3.1~\cite{Faul2009} (Wilcoxon signed-rank test, $\alpha$ = .05, power = .80, $r$ = .30) indicated that 23 participants were required, and our sample of 24 met this criterion.
Participants signed up for the study as a triad, forming eight groups in total.
We set the group size of three based on prior work, which suggested that while four is the minimum size for optimal conversation~\cite{ConversationGroupSize}, smaller groups can share and experience laughter more effectively in such a small setting~\cite{LaughterGroupSize}.
We recruited eight groups that were acquainted with each other: three consisted of family members, three were lab members, one was a group of friends, and one was composed of colleagues.
To ensure the level of closeness among the triad, we evaluated their perceived closeness with each other using Inclusion of Other in the Self Scale (IOS-scale)~\cite{IOSScale}. 
The mean score was 4.46 ($SD$ = 1.28) on a seven-point scale, where one indicated the lowest perceived closeness and seven indicated the highest. 

To prevent response bias from knowing the research background, we initially told participants that the study aimed to ``understand user behavior when watching videos in virtual environments."
Participants were compensated approximately \$10 USD Amaazon gift card after completing the 60-minute experiment.
The experimental protocol was approved by the local ethics committee of the authors' institution.

\begin{table}[htbp]
\centering
\caption{Participant Information}
\label{tab:participant_info}
\begin{tabular}{l@{\hspace{0.5em}}l@{\hspace{0.5em}}l@{\hspace{0.5em}}l@{\hspace{0.5em}}r@{\hspace{0.5em}}r}
\toprule
\textbf{Group ID} & \textbf{Relationship} & \textbf{Participant ID} & \textbf{Gender} & \textbf{Age} & \textbf{IOS-Scale}~\cite{IOSScale}\\
\midrule
G1 & Colleague & P01 & Male& 28& 5\\
 &  & P02 & Female& 26& 7\\
 &  & P03 & Female& 27& 5\\
\midrule
G2 & Friends & P04 & Male& 26& 4\\
 &  & P05 & Male& 28& 4\\
 &  & P06 & Male& 28& 5\\
\midrule
G3 & Lab Member& P07 & Male& 24& 4\\
 &  & P08 & Male& 23& 4\\
 &  & P09 & Male& 24& 5\\
\midrule
G4 & Family & P10 & Male& 63& 7\\
 &  & P11 & Female& 53& 7\\
 &  & P12 & Male& 24& 4\\
\midrule
G5 & Family & P13 & Male& 17& 5\\
 &  & P14 & Male& 56& 5\\
 &  & P15 & Male& 86& 3\\
\midrule
G6 & Family & P16 & Male& 17& 4\\
 &  & P17 & Female& 19& 5\\
 &  & P18 & Female& 57& 5\\
\midrule
G7 & Lab Member& P19 & Male& 27& 3\\
 &  & P20 & Female& 25& 2\\
 &  & P21 & Male& 42& 4\\
\midrule
G8 & Lab Member& P22 & Male& 25& 3\\
 &  & P23 & Female& 24& 3\\
 &  & P24 & Male& 26& 4\\
\bottomrule
\end{tabular}
\end{table}

\subsection{Procedure}
Upon arriving at the experimental space, participants were briefed on the experiment overview and asked to sign a consent form. 
Before the VR experience, participants completed the IOS-Scale~\cite{IOSScale} to measure their interpersonal closeness and selected their preferred avatar from the six available options.
Next, participants were instructed to sit in three separate areas of the same room for the VR session, physically separated by partitions to ensure all communication occurred exclusively through VRChat.
After being instructed to wear an HMD, participants received guidance about the virtual environment's layout and avatar control.
To facilitate avatar embodiment, we gave participants approximately 1 minute to familiarize themselves with the environment.
Participants sat on the sofa from the left side in order of their IDs and maintained the same seating positions throughout the experiment (see \autoref{fig:teaser}). 
During the study, they were informed that they could freely talk to each other through the voice chat function in VRChat.

Before watching the comedy episodes, participants watched a 2-minute practice video clip to familiarize themselves with the laughter expression control operation and communication in the virtual environment.
The main viewing session consisted of four 7-minute comedy episodes, with condition order counterbalanced across groups.
After each episode, participants removed their HMD and headphones to complete an engagement questionnaire on their smartphones.
Following the viewing sessions, participants joined a 10-minute group interview.

\subsection{Measurements}
To address our three research questions (RQ1–RQ3), we adopted a mixed methods approach combining qualitative and quantitative data sources.
This enabled us to explore both participants’ subjective experiences and observable patterns of behavior within the VR co-viewing context.

This methodological choice was intended to capture how the presence or absence of visual laughter expressions (embodied laughter cues) shaped participants' co-viewing experiences, focusing not merely on the amount of observable behavior but on how participants subjectively experienced these interactions.
RQ1 focused on individual engagement, examined through self-reported scores from a standardized engagement questionnaire and qualitative reflections from group interviews.  
RQ2 addressed emotional contagion, examined through quantitative analysis of system-logged chained laughter events and qualitative insights from group interviews.
RQ3 examined norms of emotional expression, analyzed through interview data on how participants interpreted the timing and appropriateness of laughter.


\subsubsection{Semi-structured Interview}
The semi-structured interviews were designed to explore how participants perceived and made sense of the presence or absence of embodied laughter cues during the co-viewing experience.
We asked participants how their reactions or interpretations changed when others’ laughter was visually observable, and conversely, how they were aware of others’ attention or responses when they themselves expressed laughter.
Follow-up questions were used to probe specific moments when laughter appeared to spread across participants, aiming to uncover the affective and interpretive processes underlying such chains of reaction.
All interview questions were grounded in the three predefined research questions (RQ1–RQ3) and intended to elicit subjective experiences relevant to each.
The full interview guide is provided in Appendix A.
All interviews were transcribed verbatim and analyzed using thematic analysis following the method proposed by Braun and Clarke~\cite{ThematicAnalysis}.
All authors have prior experience publishing qualitative research on HCI and VR.
We were also mindful that the lead analyst might possess a favorable bias toward VR, so we employed memos and team-based reviews to minimize interpretive bias.
The first author conducted initial open coding through close reading of the transcripts, extracting segments relevant to each RQ.
The analysis continued until thematic saturation was reached.
The naming and organization of themes were refined through discussion to ensure conceptual consistency, and the final themes were reviewed by all authors to confirm their alignment with the overarching research goals.
The complete interview guide is provided in Appendix A.

\subsubsection{Engagement: Quantitative Support for RQ1}
We adopted the questionnaire developed by~\cite{Engagement} to capture participants' engagement while watching comedy videos in VR.
While the original questionnaire included six dimensions of engagement, we focused on five dimensions most relevant to our investigation of social co-viewing experience: Focused Attention (absorption and loss of time awareness), Perceived Usability (ease of interaction and emotional response), Endurability (overall experience and willingness to return), Novelty (curiosity and interest), and Involvement (degree of being drawn into the experience). 
We excluded the aesthetics dimension in the original items because it has low relevance to our research.

There were 31 items in total, and they were rated using a 7-point Likert scale, where one indicated ``strongly disagree" and seven indicated ``strongly agree."
We revised a few words in the original questionnaire to fit the current video-viewing scenario
(e.g., \textit{``I was so involved in watching videos that I lost track of time."}).
We calculated the mean engagement score for each condition by averaging the scores from each co-viewing sessions to represent overall engagement.

\subsubsection{Emotional Contagion: Quantitative Support for RQ2}
To quantitatively support RQ2, we analyzed chained laughter events by counting how often laughter spread between participants and identifying who initiated and who followed in each sequence.
Using system logs from VRChat, we defined a chained laughter event as a sequence in which one participant triggered a laughter cue, followed within 5 seconds by a laughter cue from another participant.
This 5-second window was chosen to align with the length of the laughter animation displayed by the avatar (~4 seconds), allowing sufficient time for participants to visually perceive and react to others’ laughter cues.
If more than one participant responded within this time frame, each was counted as a separate follower.
For example, if Participant A laughed and both B and C followed within five seconds, we recorded two chains: “A → B” and “A → C.”
We aggregated these sequences to calculate how frequently each participant acted as an initiator, i.e., the first person to express laughter.


\section{RESULTS}
\begin{figure}
    \centering
    \includegraphics[width=\textwidth]{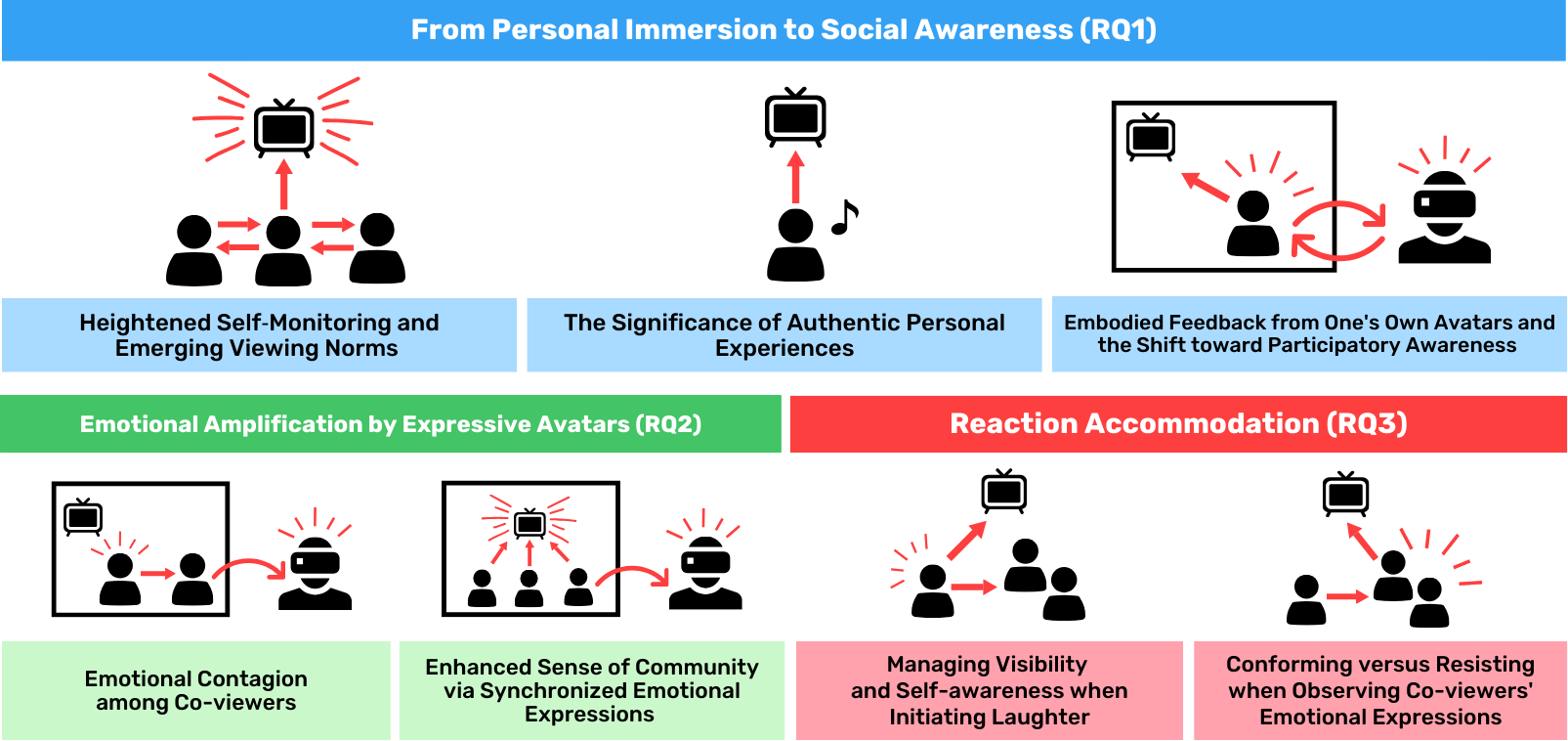}
    \caption{Themes mapped to our three research questions. First, participants’ \textbf{self‑awareness and engagement} shifted once co‑viewers’ expressions became visible (Section~\ref{sec:Content to Social Context}). Second, visibility of laughter cues promoted \textbf{emotional contagion} and mutual responsiveness (Section~\ref{sec:Emotional Amplification}). Third, groups developed \textbf{normative coordination} in how and when they displayed laughter (Section~\ref{sec:Reaction Accommodation}).}
    \label{fig:thematic}
\end{figure}

In this section, to answer our research questions, we present the qualitative findings from the semi-structured interview first (Section \ref{sec:Content to Social Context}, \ref{sec:Emotional Amplification}, and \ref{sec:Reaction Accommodation}), integrating them with relevant quantitative results to provide a comprehensive understanding of each research question.
The results of our thematic analysis are summarized in Figure~\ref{fig:thematic}, and the corresponding quantitative findings are presented in Table~\ref{tab:combined_results_rq1_rq3}.

\begin{table}[]
\small
\centering
\caption{Summary of Quantitative Results for RQ1 and RQ2}
\label{tab:combined_results_rq1_rq3}
\begin{tabular}{llccccc}
\toprule
\textbf{Category} & \textbf{Measure} & \textbf{VO (M±SD)} & \textbf{VLC (M±SD)} & \textbf{\textit{p}} & \textbf{Effect Size} & \textbf{Sig.} \\
\midrule
\multirow{5}{*}{Engagement (RQ1)} 
 & Focused Attention & 4.96 ± 0.08 & 5.03 ± 0.08 & .456 & $r = 0.05$ & No \\
 & Perceived Usability & 5.78 ± 0.07 & 5.69 ± 0.08 & .166 & $r = -0.06$ & No \\
 & Endurability & 4.88 ± 0.10 & 5.16 ± 0.09 & .014 & $r = 0.18$ & Yes \\
 & Novelty & 4.99 ± 0.11 & 5.31 ± 0.10 & .021 & $r = 0.21$ & Yes \\
 & Involvement & 4.93 ± 0.16 & 5.41 ± 0.14 & .011 & $r = 0.32$ & Yes \\
\midrule
\multirow{3}{*}{Emotional Contagion (RQ2)} 
 & Chained Laughter & 3.50 ± 4.26 & 6.38 ± 5.65 & .006 & $r = 0.87$ & Yes \\
 & Initiations & 10.5 ± 5.66 & 9.06 ± 4.76 & .210 & $d = 0.48$~\tablefootnote{Initiations was analyzed using a paired $t$-test as its normality assumption was not violated; hence, Cohen’s $d$ is reported instead of $r$.}& No\\
 & Follows & 3.31 ± 3.57 & 5.31 ± 4.88 & .072 & $r = 0.71$ & No \\
\bottomrule
\end{tabular}
\end{table}

\subsection{From Personal Immersion to Social Awareness (RQ1)}
\label{sec:Content to Social Context}
Our analysis showed how embodied laughter cues changed the viewing experience from individual content engagement to a social engagement, where viewers constantly monitored and evaluated \emph{their own} reactions in a socially negotiated context.
This shift manifested in participants' descriptions of their attention and engagement with the content versus social dynamics.

\subsubsection{Heightened Self‑Monitoring and Emerging Viewing Norms}
\label{subsec:1-1}
When seeing the expressive avatars present next to the viewing content while co-viewing, participants' attention was distributed across individual content appreciation and the co-viewing activity. 
This division of attention was reflected when P05 shared that \textit{``With avatars present, my attention was divided, with about 30\% focused on others' reactions rather than the comedy."} 
P22 further described this transformation: \textit{``With avatars, I found myself focusing less on the video itself and more on anticipating when and how to react."} 
Observing the reaction of co-viewers' avatars made participants experience social pressure and normative behaviors, as P24 explained: \textit{``When other people reacted to the comedy, I felt peer pressure to conform."} 
Here, normative behavior refers to the subtle pressure to align one’s reactions with those of the group.
This shift towards collective viewing was explicitly acknowledged, with P06 noting: \textit{``Rather than being immersed in the comedy itself, it became more about watching together as a group."}
Participants also described the specific form of social norms they perceived in these settings. P07 remarked, \textit{``When I could see other avatars, I felt like I had to empathize with them.''} This comment illustrates how the visibility of others’ expressive avatars fostered a shared expectation to display emotional resonance, even when one's own reaction might have differed. 

\subsubsection{The Significance of Authentic Personal Experiences}
\label{subsec:1-2}
In the \textbf{VO} condition, participants were able to maintain a more personal engagement with the comedy. 
Although they could hear their co-viewers' laughter through voice chat, they could not see any avatar-based expressions or movements. 
As a result, many participants described the experience as solitary despite the shared audio channel. 
For example, P05 reflected: \textit{``Even though I was participating together, it felt like no one was really there because I couldn’t see anyone."} 
The absence of social presence of co-viewers allowed participants to experience more authentic individual reactions, as P10 noted:
\textit{``When others couldn’t see me, I could be more honest with my reactions. I could laugh exactly when I personally felt something was funny, without worrying about how others might perceive it."}
Several participants recognized the value of \textbf{VO}, with P16 stating: \textit{``For the content I want to focus on or watch alone, I would prefer to view it without avatars. For example, when watching something like a lowbrow variety show, I’d rather not have others watching me."} 
These accounts highlighted how the absence of the co-viewers' social presence preserved the possibility for a more authentic, individual appreciation of the content.

\subsubsection{Embodied Feedback from Own Avatar and the Shift toward Participatory Awareness}
\label{subsec:1-3}
Participants also described how the visibility of their own avatars created a heightened awareness of being part of a shared viewing space, which shaped not only how they perceived others but also how they chose to express themselves. 
Several noted that their avatars served as embodied feedback mechanisms, reinforcing their emotional experiences and prompting deeper engagement. 
As P22 noted, \textit{``Having an avatar and seeing it clapping made me feel like I was expressing more reactions.''} 
P23 shared a similar observation, \textit{``I felt there was a positive feedback that reinforced my feeling of amusement.''}
However, the absence of expressive avatars could diminish participants' motivation to express emotions voluntarily. 
For example, in \textbf{VO} condition, P20 described, \textit{``I didn't feel motivated to use the the laughter cue trigger when there was no avatar.''} 

Our findings also revealed that the presence of avatars encouraged participants to become active contributors to the shared viewing experience by intentionally expressing their emotions. 
P07 described this proactive approach, noting, \textit{``I tried to express amusement even at small moments, thinking that my laughter might encourage others to join in.''} 
Additionally, P20 reflected on the sense of influence avatars provided, stating, \textit{``I felt more inclined to press the laughter cue trigger because being embodied in an avatar felt like I was engaging in a shared space."} 
This intentional emotional expression suggested that embodied avatars influenced participants' emotions during co-viewing, prompting them to visibly exhibit their positive feelings through the movement of the embodied avatar.

\subsubsection{Level of Engagement}
The quantitative results offer partial support for the interview findings. 
We conducted a Wilcoxon signed-rank test because the data normality assumption was violated (Shapiro-Wilk's normality test, \textit{p} < .05). 
The results of the five sub-dimensions of engagement showed that participants who watched comedy with the presence of expressive avatars had significantly higher scores in Endurability (\textit{p} = .014, Cohen's \textit{r} = 0.18), Novelty (\textit{p} = .021, Cohen's \textit{r} = 0.21), and Involvement (\textit{p} = .011, Cohen's \textit{r} = 0.32), than without seeing expressive avatars.
However, we did not find significant differences in the sub-dimension of focused attention (\textit{p} = .456, Cohen's \textit{r} = 0.05) and perceived usability (\textit{p} = .166, Cohen's \textit{r} = -0.06) between the two conditions. (\autoref{tab:combined_results_rq1_rq3})
These findings complement the qualitative accounts by suggesting that, despite divided attention and increased social awareness, the presence of expressive avatars enhanced the overall appeal and memorability of the co-viewing experience.

\subsection{Emotional Amplification by Expressive Avatars (RQ2)}
\label{sec:Emotional Amplification}
The analysis revealed how avatar-mediated co-viewing created multiple emotional amplification and contagion. 
Participants described how their emotional experiences were amplified through three distinct but interconnected mechanisms: embodied feedback from their avatars, emotional contagion between co-viewers, and enhanced social presence through synchronized emotional expressions.

\subsubsection{Emotional Contagion among Co-viewers}
\label{subsec:2-1}
In addition to amplifying one's emotion through the expressive avatars, we found that emotions were amplified through collective experiences and emotional contagion, where participants’ emotional responses were often triggered or intensified by co-viewers’ reactions. 
P07 described this phenomenon, saying, \textit{``When I noticed someone pressing the laughter cue trigger through their movement, I strongly felt the chain reaction of laughter. About half of my laughter came from the fact that everyone else was laughing.''} 
This contagious effect was explicitly acknowledged by P23: \textit{``When everyone else was pressing the laughter cue trigger, I started to find it a bit amusing too.''} 
P04 captured how such contagion could resolve hesitation, stating, \textit{``When others laughed, I laughed along. When I was unsure, seeing them made me feel it was okay to trigger the button.''} 

\subsubsection{Enhanced Sense of Community via Synchronized Emotional Expressions}
\label{subsec:2-2}
\paragraph{Chained Laughter.}
We conducted a Wilcoxon signed-rank test because the data normality assumption was violated (Shapiro-Wilk's normality test, \textit{p} < .05). 
The result showed that when watching comedy with expressive avatars, participants had significantly higher chained laughter than without seeing expressive avatars ($V$ = 0, \textit{p} = .006, Cohen's \textit{r} = .87, \autoref{tab:combined_results_rq1_rq3}).
This result was in line with the qualitative finding in Section \ref{subsec:2-1}, suggesting that the presence of expressive avatars facilitated positive emotional contagion when co-viewing in social VR.

\paragraph{Initiating and Followed Laughter.}
A Shapiro-Wilk normality test indicated that the difference scores for initiating laughter met the normality assumption ($W$ = .902, \textit{p} = .299), allowing us to conduct a paired \textit{t}-test. The result showed no significant difference between conditions (\textit{t}(7) = 1.35, \textit{p} = .220), although the effect size was medium (Cohen’s \textit{d} = .48). For followed laughter, the normality assumption was violated ($W$ = .744, \textit{p} = .007), and a Wilcoxon signed-rank test showed a marginally significant difference, with more followed laughter observed in the expressive avatar condition ($V$ = 2, \textit{p} = .072, \textit{r} = .71, \autoref{tab:combined_results_rq1_rq3}).

Participants reported that seeing embodied laughter cues from others made it easier to feel that they were reacting to the same moments together. 
This perceived synchronization of laughter enhanced their sense of community and helped transform the viewing experience into a collective activity. 
P09 stating, \textit{``When we found the same moments amusing, I strongly felt the presence of others.''} Avatar-mediated emotional synchrony provided a sense of shared space and co-presence, as P01 shared: \textit{``Having avatars made it feel like we were watching together.''} 
P23 highlighted the impact of shared timing, noting, \textit{``When everyone pressed the button at the same time, it made the moment feel even funnier.''}
This collective experience indicated that expressive avatars enhanced the interactive co-viewing experience with remote viewers. The embodied avatars not only made people feel more engaged while consuming the video but also strengthened the emotional bonds among remote co-viewers.

\subsection{Reaction Accommodation (RQ3)}
\label{sec:Reaction Accommodation}
We found that participants adjusted their emotional expressions in response to others, which was potentially influenced by the avatars’ physical presence. 
We found that participants managed their presentation of emotional expressions when initiating laughter and following co-viewers' emotional cues via the embodied avatar.

\subsubsection{Managing Visibility and Self-Awareness when Initiating Laughter}
\label{subsec:3-1}
When participants initiated emotional reactions at humorous moments, they became highly aware of co-viewers' presence, carefully managing their expressions to align with the perceived group atmosphere or change the group atmosphere.  
P09's comment explained this self-awareness of co-viewers' reaction, \textit{``When I find something funny, I tend to pay attention to whether others find it funny too."}
They further noted that \textit{When we find the same moments amusing, I feel reassured.''}
These comments suggest that laughter was not only a personal response but also socially regulated: participants found comfort when their reactions aligned with those of the group, and monitored others to ensure that their expressions fit the perceived shared experience.
P23 elaborated on the impact of visibility of their reaction through avatars, noting, \textit{``I became very conscious that others could see my reactions through my avatar, so I felt pressure to consider how they would interpret my amusement. There were times that I was the first in the group to burst out laughing, and I felt pressure about thinking of how others might interpret my amusement.''}  
This comment suggests that such pressure was not simply due to being heard, but rather emerged from the visual presence of avatars that made emotional expressions publicly observable, unlike in voice-only conditions.

\subsubsection{Conforming versus Resisting when Observing Co-viewers' Emotional Expressions}
\label{subsec:3-2}
We found that participants made conscious decisions about whether to conform to or diverge from co-viewers' emotional expressions when sensing their reactions through the expressive avatars.  
When observing co-viewers reacting to the comedy, some participants said they \textit{``would look at others' reactions first, and then realize 'Oh, that was funny' before pressing [the laughter cue trigger]''} (P18).
P04 added, \textit{``Seeing others [avatar] laughing reassured me that it was okay to press [the laughter cue trigger] when I was wondering if I should.''} 

However, asking participants to actively indicate a humorous moment and presenting them with expressive avatars can lead to excessive focus on the co-viewing situation, as they may become preoccupied with the reactions of their co-viewers and the content.
As P01 shared, \textit{``When seeing other people laughing and I didn't, I found myself wondering about their intentions or what they found funny."} This participant also highlighted instances of resisting conformity: \textit{``[When I saw others laughing] I found it somewhat amusing as well, but not enough to press the [the laughter] button.''}  
P08 shared, \textit{``I would observe others, thinking `Someone's probably going to laugh here... ah, there it is.'"}  
P10 noted the perceived difference in intensity of expressions, saying, \textit{``Since I was only expressing slight amusement, I felt that others might be expressing more intense laughter."}

These comments revealed that participants intentionally pressed the laughter cue trigger to control the expressive avatars by accommodating or not accommodating co-viewers' reactions.

\section{DISCUSSION}
\subsection{Reframing Engagement through Expressive Avatars (RQ1)}
\subsubsection{Balancing Social Engagement and Personal Immersion}
Our findings suggest that the introduction of embodied laughter cues shifted users’ engagement from solitary immersion to heightened social awareness.
In Section \ref{subsec:1-1}, we found that participants became increasingly attentive to how their co-viewers reacted during co-viewing when being embodied in expressive avatars.
This social awareness led to a split attention between the comedy content and others’ reactions.
Despite participants dividing their attention between comedy and their embodied co-viewing partners, we still found significantly increased Involvement, Endurability, and Novelty when co-viewing with expressive avatars.
These dimensions respectively reflect the depth of experiential immersion, willingness to re-engage, and curiosity toward the interactive aspects of the system.
Taken together, our findings indicate that, in the context of co-viewing comedy with close social ties, embodied laughter cues generally enhanced the co-viewing experience.
These results challenge the assumption that social distractions inherently detract from engagement~\cite{AvatarCollaborative}, indicating that they could actually foster different kinds of socially mediated involvement in co-viewing settings.

However, the finding in Section \ref{subsec:1-2} emphasized the equal importance of maintaining personal immersion during co-viewing.
In the \textbf{VO} condition, participants heard co-viewers' laughter through voice chat but were not visually observed.
Many participants found this setting to be more conducive to honest reactions, as they felt less observed or judged when embodied in expressive avatars.
Some noted that the absence of visible co-viewers made it easier to focus on the content and laugh freely when they genuinely felt amused.
Others preferred this mode for content they wanted to enjoy alone, such as niche or controversial forms of comedy, because it allowed for a sense of privacy even while co-viewing.


\subsubsection{Enhancing Engagement through Feedback Loops from Self-Avatars}
The result showed that the embodied laughter cues expressed through participants’ own avatars formed a feedback loop that enhanced their engagement by promoting a sense of active involvement (Section \ref{subsec:1-3}).
This mechanism aligns with the Facial Feedback Hypothesis~\cite{FacialFeedbackHypothesis}, which posits that producing facial expressions such as smiling can elicit positive emotional states.
In particular, the visual reinforcement of one’s own smile through avatar representation appeared to intensify affective responses, consistent with prior studies involving mirror-based systems~\cite{SmileShigeo} that display algorithmically transformed smiling faces of users to induce positive emotions.
These studies have shown that viewing one's face transformed into a smile can amplify positive feelings.

Moreover, the visualization and recursive observation of one’s emotional expression through the embodied avatar can be interpreted in relation to an expanded understanding of the Proteus effect~\cite{ProteusEffect}.
While the Proteus effect originally referred to behavioral change induced by the stereotypical associations of avatar appearance, recent discussions have considered the broader influence of avatar-based self-representations on users’ sensory and affective experiences.
In this context, the feedback loop of laughter expression observed in our study can be positioned as an instance in which an embodied avatar modulates the perception of one’s own emotional state.
These findings suggest that embodied self-representation not only externalizes users’ emotional expressions but also modulates their internal affective experience, thereby deepening their engagement with the co-viewing activity itself.
These insights suggest that co-viewing systems may benefit from considering how expressive self-avatars support emotional feedback loops that shape user engagement.


\subsection{Emotional Contagion and Amplification through Expressive Avatars in Co-Viewing (RQ2)}

As shown in Section \ref{sec:Emotional Amplification}, embodied laughter cues shaped participants’ emotional experiences in co-viewing by facilitating mutual responsiveness and collective alignment.
We identified two themes that characterized how this occurred.
\textbf{Emotional Contagion among Co-viewers} (Section~\ref{subsec:2-1}) captures that participants reported laughing when they visually observed laughter responses from their co-viewers.
For example, P07 shared, \textit{“About half of my laughter came from the fact that everyone else was laughing.”} Such comments indicate that laughter was not always internally generated, but sometimes followed visible expressions from co-viewers.
 Although the laughter cues were manually triggered, participants often acted in response to others’ reactions rather than solely their own feelings. This aligns with prior findings in text-based co-viewing, where emotional contagion was inferred from sentiment alignment across distributed comments~\cite{DanmakuIntJHumComputInteract, EmotionalContagionInCMC}.
 Our study extends these insights to embodied VR, where expressive avatars made emotional responses more visible and temporally aligned. Just as participants experienced emotional reinforcement through self-avatar feedback, the visible reactions of others also appeared to trigger affective responses. This socially guided behavior aligns with theoretical accounts of emotional contagion grounded in the mirror neuron system~\cite{MirrorNeuron,MirrorNeuron2,EmotionContagion}, which facilitates unconscious mimicry of others’ bodily expressions. Because avatar-mediated cues are visually embodied, such mechanisms may operate more strongly in VR than in text-based communication. While further investigation is needed, our findings point to the potential of embodied systems to elicit affective responses through neural and behavioral mimicry.

This process of emotionally responsive behavior also connects to our second theme, \textbf{Enhanced Sense of Community via Synchronized Emotional Expressions} (Section~\ref{subsec:2-2}), as participants sometimes described a stronger feeling of togetherness when they laughed at the same moments as others.
While our quantitative analysis did not show significant differences in the frequency of laughter initiations or follows, this may reflect that participants could already hear each other's laughter in the voice-only condition.
Nonetheless, interview responses suggest that expressive avatars helped participants interpret and coordinate emotional expressions more intentionally.
Rather than increasing how often people laughed, the system supported subtle emotional tuning within familiar groups.
From a design perspective, expressive avatars may be especially effective in co-viewing scenarios among close social ties, where lightweight emotional alignment can enrich the shared experience without requiring explicit synchronization.



\subsection{Accommodating Emotional Expressions in Avatar-Mediated Co-Viewing (RQ3)}
\subsubsection{Accommodate Strategies through Expressive Avatars}
We observed that participants adjusted their emotional expressions in response to co-viewers when engaging in VR co-viewing with expressive avatars.
Specifically, such adjustments were evident in two contexts: when initiating laughter themselves (Section~\ref{subsec:3-1}), and when responding to laughter initiated by others (Section~\ref{subsec:3-2}).
This behavioral pattern aligns with Communication Accommodation Theory (CAT)~\cite{CAT,CAT2}, which describes how individuals adjust their communication to align (or not align) with their conversational partner based on various communication and relational goals.
It has been found that people could accommodate to virtual agent which represented in an avatar at the language level in a virtual interview context~\cite{QuidProQuo}. In line with this research, our findings extend further by showing that people accommodate to each other at affective level in avatar-mediated interaction, where nonverbal behaviors are embodied and rendered via expressive avatars.

In \ref{subsec:3-1}, we observed a pattern consistent with CAT terms an accomodative orientation.
Participants adjusted the form and timing of their emotional expressions before receiving clear feedback from others. Their comments suggest that this regulation was motivated by a desire to avoid standing out and to harmonize with the group’s expected emotional tone.
In \ref{subsec:3-2}, participants made situated decisions about whether to join others in laughter or to refrain.
This behaviour entails both reactive convergence, in which participants echoed others’ laughter for reassurance, and divergence, in which they deliberately withheld or softened their reaction to preserve personal boundaries.
While prior computer-mediated communication (CMC) studies have shown that linguistic style matching and emoticon use promote group cohesion~\cite{EmoticonConvergenceCMC,LanguageStyleMatchingCMC}, and synchronization with non-human avatar agents can enhance feelings of affiliation~\cite{FUJIWARA2022107079}, empirical understanding of how embodied humans mutually adjust nonverbal behavior in VR has been limited.
Our study provides initial understanding to contribute to this gap by showing how expressive avatars mediate nonverbal accommodation  in immersive settings, thereby extending the theoretical scope of CAT to embodied, avatar-mediated communication.

\subsubsection{Design Considerations for Supporting Emerging Co-viewing Norms}
These findings inform key avatar design considerations for establishing viewing norms that balance engagement and peer-driven emotional contagion and amplification in VR co-viewing.
First, the expressiveness of avatars should correspond to the clarity of the viewing norm. For content where emotional responses are commonly shared, such as comedy, live sports events, or horror films with synchronized pacing, enabling rich embodied cues can facilitate mutual responsiveness. This approach may help reduce ambiguity about when or how to react. It is particularly effective when viewers are expected to experience similar levels of emotional intensity.
Second, users should be able to regulate the visibility of others' emotional expressions to match their own comfort and attention preferences. Our findings indicate that high visibility can support affective alignment, but may also introduce pressure to conform.
In Social VR platforms such as VRChat, users can choose to display only the avatars of their friends or to hide other users entirely. 
Building on this, systems could allow users to have autonomy to control which avatars are visible and how they are rendered. This kind of selective control can help balance social awareness with individual control. 
Together, these principles suggest that co-viewing norms in avatar-mediated interaction emerge through ongoing coordination among participants, rather than being predetermined. Supporting this coordination through flexible and context-sensitive system features can enhance both individual engagement and group-level emotional coherence.

\section{LIMITATION AND FUTURE DIRECTIONS}
This study deliberately focused on laughter as a specific category of emotional expression and on comedy as the viewing genre.
This scope was chosen to ground our investigation in a well-established affective signal that frequently appears in co-viewing settings, particularly in shared commentary systems such as danmaku~\cite{LaughterComedy1}.
Nonetheless, laughter represents only one facet of the broad spectrum of human affect. 
Building on the current finding, future research could investigate how people experience emotional contagion with co-viewers for various emotional expressions, such as fear, sadness, or surprise, while being embodied in expressive avatars during co-viewing of different genres.
Given the challenges of conveying subtle nonverbal cues in social VR~\cite{SocialVRNonverbal1,SocialVRNonverbal2}, extending this research to other emotions and genres would require further development in avatar expressivity as well as integration with lightweight affect sensing technologies.
In the current study, we employed a button-based input mechanism to ensure participants have full autonomy in controlling the expression of their embodied avatars.
Future systems could incorporate wearable sensing to support more spontaneous and multimodal emotional interaction, broadening the communicative bandwidth available in co-viewing scenarios.

This study also centered on small groups of familiar participants, reflecting the practice that verbal co-viewing is often conducted among those who already knew each other in social VR \cite{SmallGroup1,SmallGroup2}.
Communication Accommodation Theory (CAT) suggests that convergence and divergence patterns differ markedly depending on group familiarity and social relationships~\cite{CAT}.
Social VR platforms enable people to share experiences with larger, more heterogeneous groups, such as during virtual concerts or sports events.
Though our study showed that the presence of expressive avatars had an effect on co-viewing with family and friends, it is also interesting to further explore how being embodied in expressive avatars influences co-viewing experiences with strangers in a large group, such as co-viewing large online events in social VR.
Understanding how strangers create emotional convergence through embodied avatars offers valuable insights into the scalability of avatar-mediated emotional expression. 

\section{CONCLUSION}
This study examined how the introduction of embodied laughter cues, as compared to voice-only co-viewing, influences engagement, emotional contagion, and social interaction during co-viewing experiences in virtual reality.
The results showed that embodied laughter cues led to a shift in the nature of engagement, from personal immersion to collective and norm-driven engagement, and increased several engagement-related measures.
The presence of expressive avatars also facilitated shared emotional experiences, as participants became more aware of their own and co-viewers' reactions, leading to positively valenced emotional contagion.
While this shift introduced a trade-off between content focus and social attention, it enabled the emergence of negotiated viewing norms that supported fluid forms of participation.
Future work should examine how these findings apply to content that evokes more complex emotional responses and to varied group configurations.  
In particular, future systems should explore how to support the co-construction of viewing norms across diverse settings through flexible and adaptive control of emotional visibility.
 

\bibliographystyle{ACM-Reference-Format}
\bibliography{main_ref}

\appendix
\section{Interview Guide}
\label{appendix:interview}

The semi-structured interviews covered the following topics and questions:
\begin{itemize}
   \item Social Presence
       \begin{itemize}
           \item During the study, how did you feel others' presence?
           \item How did the presence of avatars affect your viewing experience?
       \end{itemize}
   \item Nonverbal Communication
       \begin{itemize}
           \item When seeing the avatars present above the TV screen, how did it influence your feelings about other co-viewers' reactions?
           \item Have you ever tried to compensate for the lack of nonverbal cues when avatars were absent? If yes, how?
       \end{itemize}
   \item Verbal Communication
       \begin{itemize}
           \item How did the presence of avatars influence your communication while watching TV in VR?
           \item How did you feel about the interaction with each other or the depth of the conversation when the avatar was present and absent?
       \end{itemize}
   \item Laughter Button Usage
       \begin{itemize}
           \item How did the use of the \textit{laughter button} differ between the two conditions?
           \item How did you notice (or not notice) other co-viewers' \textit{laughter button} usage when avatars were present?
       \end{itemize}
   \item Potential Applications
       \begin{itemize}
           \item Based on your experience so far, what types of scenarios would you like to use this system?
           \item What other types of videos would you like to watch with avatars?
       \end{itemize}
\end{itemize}









\end{document}